\documentclass[twocolumn,3p,times]{elsarticle}

\usepackage{amssymb}
\usepackage{graphics}
\usepackage{graphicx}
\usepackage{txfonts}
\usepackage{epsfig}
\usepackage{bm}   
\usepackage{rotate}
\usepackage{amsmath,amsfonts}
\usepackage{longtable}
\usepackage{latexsym,rotating,wrapfig}
\usepackage{footmisc,booktabs,amssymb,longtable}

\usepackage{xcolor,colortbl}

\usepackage{multirow}
\definecolor{Gray}{gray}{0.85}
\definecolor{LightCyan}{rgb}{0.88,1,1}

\usepackage[scaled=0.92]{helvet}
\usepackage[T1]{fontenc}
\usepackage{textcomp}
\usepackage{braket}
\usepackage{textgreek}
\usepackage{rotate}
\usepackage{tabularx}
\usepackage[referable]{threeparttablex}
\journal{Physics Letters B}

\begin{document}

\begin{frontmatter}

\title{Triaxial nuclear shapes from simple ratios of electric-quadrupole matrix elements}

\author[a,b]{Elena Atanassova Lawrie}
\ead{ea.lawrie@ilabs.nrf.ac.za}

\author[b,c]{Jos\'e Nicol\'as Orce}
\ead{jnorce@uwc.ac.za}
\ead[url]{https://nuclear.uwc.ac.za}

\address[a]{iThemba LABS, National Research Foundation, PO Box 722, Somerset West 7129, South Africa}
\address[b]{National Institute for Theoretical and Computational Sciences (NITheCS), South Africa}
\address[c]{Department of Physics \& Astronomy, University of the Western Cape, P/B X17, Bellville ZA-7535, South Africa}

\date{\today}

\begin{abstract}

Theoretical models often invoke triaxial nuclear shapes to explain elusive collective phenomena, but such assumptions are usually difficult to confirm experimentally.
The only direct measurements of the nuclear axial asymmetry $\gamma$ is based on rotational invariants of zero-coupled products of the electric-quadrupole ({\sc E2}) operator,
the Kumar-Cline sum rule analysis, which generally require knowledge of a large number of {\sc E2} matrix elements connecting the state of interest.
We propose an alternative assumptions-free method to determine $\gamma$ of even-even rotating nuclei using only two {\sc E2} matrix elements,
which are among the easiest to measure.
This approach is based on a simple description of nuclear rotation, where the underlying assumptions of the Davydov-Filippov model are either empirically proven or unnecessary.
The $\gamma$ values extracted here are found in  agreement with the values deduced from Kumar-Cline sum rules measurements (where available), providing further evidence that the proposed approach represents a reliable, model-independent deduction of $\gamma$.
The technique was applied to more than 60 deformed even-even rotating nuclei and the results indicate that rotating nuclei generally exhibit well-defined stable axially-asymmetric shapes.

\end{abstract}

\end{frontmatter}


Triaxial shapes  --- like kiwis or flattened footballs ---  break the axial symmetry of a deformed object and are basic ingredients in theoretical models describing both the quantum world and the realm of general relativity, albeit its testing
through direct experimental observations remains challenging. Triaxiality plays an important role in (i) nuclear fission~\cite{afanasjev2013nuclear},
with its relevance to energy production; (ii)  the radiative capture of neutrons in stellar explosions~\cite{grosse2014broken}, responsible for the creation of heavy elements; (iii) the formation of some superdeformed bands in nuclei~\cite{riley2009strongly},  and iv) the low-lying nuclear structure~\cite{garrett2019multiple}.

The majority of  nuclei 
show quadrupole deformations~\cite{1998nuclear,rowe2010fundamentals}, 
described 
by two parameters, $\beta_{_2}$ and $\gamma$. 
Here, $\beta_{_2}$ defines the magnitude of the quadrupole  deformation and $\gamma$ the degree of axial asymmetry or triaxiality, where axially-symmetric deformations correspond to $\gamma$ = 0$^\circ$ (prolate) and $\gamma$ = 60$^\circ$ (oblate) while triaxial shapes to $0^\circ < \gamma < 60^\circ$. 
Global calculations of all even-even nuclei in the nuclear chart~\cite{moller2006global,moller2008axial} suggest that the total energy of many nuclei decreases substantially
if the nuclear shape has stable triaxial deformation.

Theoretical approaches where the $\gamma$ degree of freedom plays a dominant role involve $\gamma$ vibrations and rotations.
The former may appear as (i) a dynamical feature of the nuclear shape, corresponding to small $\gamma$ oscillations of the nuclear surface around an average axially symmetric shape~\cite{bohr1952coupling}, and (ii) as large-scale $\gamma$ oscillations caused by the $\gamma$-softness of the nuclear shape, that may cover the whole range of $\gamma$ between 0$^\circ$ and 60$^\circ$~\cite{wilets1956surface}.
In contrast, deformed nuclei with stable triaxial shape rotate around their three axes generating sets of rotational bands that 
can be described within the Davydov-Filippov (DF) model~\cite{davydov1958rotational,davydov1959relative,hecht1962asymmetric}. This rotation looks like the precession of a rotating top.

Triaxial deformation has often been inferred through indirect methods by comparing experimental observations with the predictions of theoretical models,
based on (i) the splitting of the giant dipole resonance ({\sc GDR}) into three dominant peaks~\cite{rezwani1970dynamic,rezwani1972further,bortolani1991isospin,mennana2021giant}, (ii) the signature splitting and inversion in rotational bands~\cite{bengtsson1984signature,ding2021signature,zeidan2002rotational,seiffert1993band}, (iii) the near-degeneracy of chiral partner bands~\cite{frauendorf1997tilted, lawrie2010reaching,shirinda2016multiple}, and (iv) the features of the tilted precession and wobbling bands~\cite{14FraPRC89,lawrie2020tilted}. Alternatively,
$\beta_2$ and $\gamma$ can be  extracted from potential energy surface calculations, 
e.g., total Routhian surface \cite{xu2000quadrupole}, Cranked Nilsson-Strutinsky \cite{carlsson2006many},
and beyond mean-field calculations of total energy surfaces and collective wave functions~\cite{orce2006shape,marchini2023emergence,garrett2019multiple}.

Rotational invariants represented as Kumar-Cline ({\sc KC}) sums~\cite{kumar1972intrinsic,cline1986heavy} remain to date the only direct experimental technique to establish the magnitude of triaxiality,
$\gamma_{_{KC}}$, in the intrinsic frame of the nucleus. Such an analysis requires experimental data on a large number of electric-quadrupole ({\sc E2}) matrix elements
of up to sixth-order {\sc E2} invariants to evaluate also statistical fluctuations, which are hard to determine experimentally.
Among more than 270 deformed rotating even-even nuclei with a ratio of excitation energies between the first $4^+_{_1}$ and $2^+_{_1}$ states of $R_{_{4/2}}\geq 2.4$~\cite{pritychenko2012national},
$\gamma_{_{KC}}$ values have only been determined for 19; namely,  $^{74,76}$Ge~\cite{toh2000coulomb,ayangeakaa2019evidence},
$^{76}$Kr~\cite{clement2007shape},
$^{98}$Sr~\cite{clement2016spectroscopic},
$^{104}$Ru \cite{srebrny2006experimental},
$^{106-110}$Pd \cite{svensson1995multiphonon,svensson1989coulomb},
$^{148}$Nd \cite{ibbotson1997quadrupole},
$^{166,168}$Er \cite{fahlander1992triaxiality,kotlinski1990coulomb},
$^{172}$Yb \cite{fahlander1992coulomb},
$^{182,184}$W \cite{wu1991electromagnetic},
$^{186-192}$Os \cite{wu1996quadrupole}
and $^{194}$Pt \cite{wu1996quadrupole}. 
These are all stable nuclei, except for $^{76}$Kr~\cite{clement2007shape} with a half-life of 14.8 h, and $^{98}$Sr~\cite{clement2016spectroscopic} with a half-life of 0.653 s,
where the corresponding {\sc E2} matrix elements were primarily extracted from multi-step Coulomb-excitation measurements~\cite{alder1960theory}.
The deduced $\gamma_{_{KC}}$ values indicate that all these nuclei present triaxial deformations,
which highlights the need for establishing a simpler model-independent approach for evaluating triaxiality.

Recently, an assumption-free approach was proposed for 
even-even rotating nuclei through the generalized triaxial-rotor model ({\sc TR}) with independent electric quadrupole and inertia tensors~\cite{wood2004triaxial}.
This approach is based on the DF model, 
but the moments of inertia (MoI) asymmetry is 
described through a new parameter $\Gamma$,
in an independent way from
the shape asymmetry $\gamma$.
Thus the assumption of the DF model that the MoI follow the irrotational-flow dependence with respect to $\gamma$ become redundant.
The generalized TR model was then applied for the 2$^+_{_1}$ and 2$^+_{_\gamma}$ states of 26 even-even rotating nuclei with $R_{_{4/2}}\geq 2.4$ \cite{allmond2017empirical,23All-CWAN},
for which experimental data on the required four  {\sc E2} matrix elements  were available.
As it was applied to the 2$^+_{_1}$ and 2$^+_{_\gamma}$ states only, it made the additional assumption of the DF model regarding the spin-dependence of the MoI also redundant;
hence, providing an empirical, assumptions-free determination of the $\gamma$ deformation for these nuclei.
Moreover, all these nuclei were found to possess triaxial deformations, supporting 
the consideration that triaxiality might be a common feature for nuclei.

In this Letter, we propose to expand this approach and 
determine the magnitude of the nuclear triaxiality of even-even rotating nuclei in the same assumption-free approach, but using only two {\sc E2} matrix elements. 
The number of required {\sc E2} matrix elements is reduced because we adopt the irrotational-flow dependence between the parameters $\Gamma$ and $\gamma$.
We consider that this dependence was proved within the assumptions-free generalized {\sc TR} approach for 12 even-even nuclei~\cite{allmond2017empirical} 
and for 13 
more even-even rotating nuclei, discussed in this work. 
Thus, the proposed analysis for the 2$^+_{_1}$ and 2$^+_{_\gamma}$ states, 
while remaining based on the {\sc DF} equations allows us to determine the $\gamma$ deformation of more than 60 even-even rotating nuclei in a simple, model-independent evaluation.


Deformed nuclei can easily be recognised by their large $B(E2; 2^+_{_1} \rightarrow 0^+_{_1})$ reduced transition probabilities values (of $\gtrapprox 20$ Weisskopf units)
connecting the first-excited 2$^+_{_1}$ and the ground 0$^+_{_1}$ states with an {\sc E2} transition.
The $B(E2)$ values for rotating nuclei are directly proportional to the square of the intrinsic quadrupole moment of the nucleus, $Q_{_0}$, and the corresponding $\langle I_{_1}~K~2~0~|~I_{_2}K \rangle$ Clebsch–Gordan coefficient~\cite{1998nuclear,ring2004nuclear}, and for axially-symmetric deformed nuclei, 
\begin{equation}
B(E2; 0^+_{_1}\rightarrow 2^+_{_1}) = 
\frac{5}{16\pi}~Q_{_0}^2,
\label{eq:rotmodel}
\end{equation}
where $Q_{_0}$ is related to $\beta_{_2}$ by~\cite{1998nuclear}
\begin{equation}
Q_{_0}=\frac{3}{\sqrt{5\pi}}~Ze R^2 \beta_{_2} \left[ 1 + 0.16 \beta_{_2}  \right],
\label{eq:beta}
\end{equation}
with $Z$ being the proton number, $R=1.2~ A^{1/3}$ fm the radius of a nucleus with a sharp surface, and $A=N+Z$ the atomic mass number with $N$ the number of neutrons.

The Hamiltonian of a deformed rotating nucleus with stable triaxial deformation comprises simultaneous rotations around the nuclear axes, 
\begin{equation}
     \mathcal{H} = \frac{\hbar^2}{2\Im_{_1}} \hat{I_{_1}^2} + \frac{\hbar^2}{2\Im_{_2}} \hat{I_{_2}^2} + \frac{\hbar^2}{2\Im_{_3}} \hat{I_{_3}^2},
\end{equation}
where $\hat{I_{_k}}$ are the operators of the total angular momentum projections onto the body-fixed axes, and $\Im_{_1}$, $\Im_{_2}$, and $\Im_{_3}$  the corresponding MoI.
The {\sc DF} model generally adopts two main assumptions about the nuclear rotation. Firstly,
the relative ratios of $\Im_{_1}$, $\Im_{_2}$, and $\Im_{_3}$ for a given $\gamma$ deformation follow the irrotational-flow dependence,
\begin{equation}
\Im_{k}(\gamma) = \Im _{_0}
\sin^2 \left( \gamma - k\frac{2\pi}{3} \right ),
\label{irro1}
\end{equation}
with $\Im_{_0}$ the MoI of an axially symmetric nucleus with respect to an axis that is orthogonal to the axis of symmetry~\cite{rowe1970nuclear}, and $k=1,2,3$.
In fact, the $\gamma$ dependence in Eq.~\ref{irro1} is more general than the irrotational-flow model (for details see Ref.~\cite{rowe2010fundamentals}, page 121).
Secondly, the {\sc DF} model needs an assumption about the spin dependence of the MoI. In the original {\sc DF}  model, the MoI remains constant as a function of spin,
whereas in later applications 
variable moments of inertia~\cite{toki1975asymmetric} are often introduced.

\begin{figure}[!ht]
\begin{center}
\includegraphics[width=7cm,height=6cm,angle=-0]{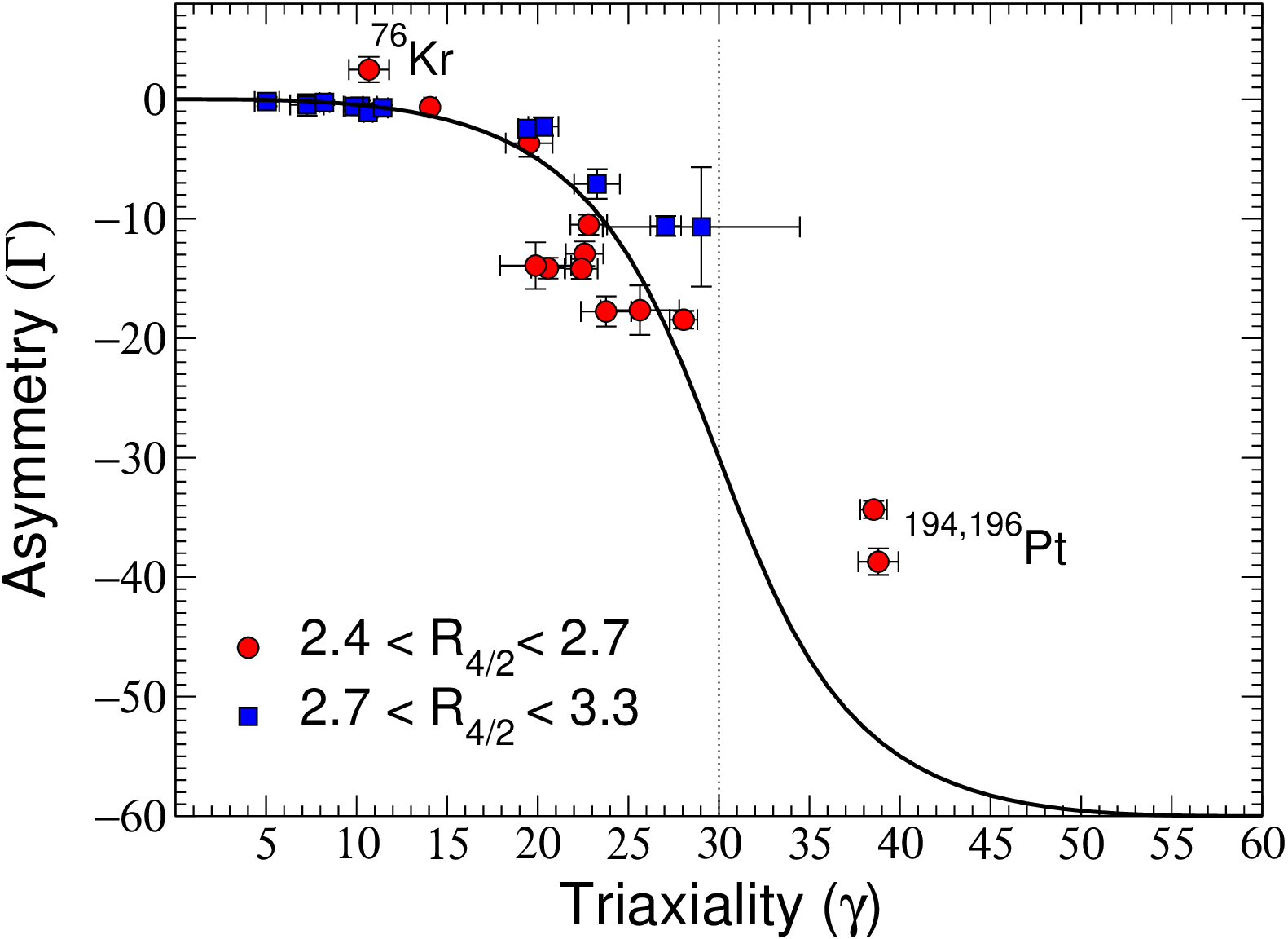}
\caption{The irrotational-flow model $\Gamma_{irr}(\gamma)$ (solid line) in comparison with empirical $\gamma_{_{TR}}$ and $\Gamma_{_{TR}}$ values for deformed even-even
nuclei with $R_{_{4/2}}$ values between 2.4 and 3.3.}
\label{fig-Gamma-vs-gamma}
\end{center}
\end{figure}

Instead of adopting the $\gamma$ dependence of  Eq.~\ref{irro1}, the generalized TR model describes the asymmetry in the three MoI independently from $\gamma$, by introducing a new MoI-asymmetry parameter $\Gamma$~\cite{wood2004triaxial}.
Accordingly, the {\sc E2} matrix elements connecting the 0$^+_1$, 2$^+_{_1}$ and 2$^+_{_\gamma}$ states are given by
\begin{eqnarray}
\langle0^+_{_1} \parallel \hat{E2} \parallel 2^+_{_1}\rangle &=& \sqrt{\frac{5}{16\pi}} ~Q_{_0} \cos(\gamma + \Gamma)\label{ME01},\\
\langle2^+_{_1} \parallel \hat{E2} \parallel 2^+_{_1}\rangle &=& -\sqrt{\frac{25}{56\pi}} ~Q_{_0} \cos(\gamma - 2\Gamma)\label{ME21} \\
&=& -\langle2^+_{_\gamma} \parallel \hat{E2} \parallel 2^+_{_\gamma}\rangle, \nonumber \\
\langle2^+_{_1} \parallel\hat{E2}\parallel 2^+_{_\gamma}\rangle &=& \sqrt{\frac{25}{56\pi}} ~Q_{_0} \sin(\gamma -2\Gamma)\label{ME22},\\
\langle0^+_{_1} \parallel \hat{E2} \parallel 2^+_{_\gamma}\rangle &=& \sqrt{\frac{5}{16\pi}} ~Q_{_0} \sin(\gamma + \Gamma)\label{ME02}.
\end{eqnarray}
Therefore, empirical values for the axial asymmetry of the shape of rotating nuclei ($\gamma_{_{TR}}$) and of the MoI ($\Gamma_{_{TR}}$) can be extracted from Eqs. \ref{ME01}, \ref{ME21}, \ref{ME22}, and \ref{ME02}~\cite{allmond2017empirical},
\begin{equation}
 \gamma_{_{TR}} = \frac{1}{3} \left[ 2 \tan^{-1}\left(\frac{\langle0^+_{_1} \parallel\hat{E2}\parallel 2^+_{_\gamma}\rangle}{\langle0^+_{_1} \parallel\hat{E2}\parallel 2^+_{_1}\rangle}\right)
+  \tan^{-1}\left(-\frac{\langle2^+_{_1} \parallel\hat{E2}\parallel 2^+_{_\gamma}\rangle}{\langle2^+_{_1} \parallel\hat{E2}\parallel 2^+_{_1}\rangle}\right) \right],
\label{gammaTR}
\end{equation}
\begin{equation}
  \Gamma_{_{TR}} = \frac{1}{3} \left[ \tan^{-1}\left(\frac{\langle0^+_{_1} \parallel\hat{E2}\parallel 2^+_{_\gamma}\rangle}{\langle0^+_{_1} \parallel\hat{E2}\parallel 2^+_{_1}\rangle} \right)
-  \tan^{-1}\left(-\frac{\langle2^+_{_1} \parallel\hat{E2}\parallel 2^+_{_\gamma}\rangle}{\langle2^+_{_1} \parallel\hat{E2}\parallel 2^+_{_1}\rangle}\right) \right],
\label{GammaTR}
\end{equation}
\noindent
using four measured {\sc E2} matrix elements. 
Note that these equations remain assumptions-free 
for rotating nuclei because they are applied to the 2$^+$ states only; i.e., the spin-dependence of the MoI becomes irrelevant.
Available experimental data allowed $\gamma_{_{TR}}$ and $\Gamma_{_{TR}}$ values to be deduced for 26 even-even rotating nuclei~\cite{allmond2017empirical,23All-CWAN},
further testing whether $\Gamma$ and $\gamma$ are independent or may follow the irrotational-flow model dependence,
 \begin{equation}
    \Gamma(\gamma) = -\frac{1}{2} \cos^{-1} \left( \frac{\cos{(4\gamma)} + 2\cos{(2\gamma)}}{\sqrt{9-8\sin^2{(3\gamma)}}} \right).
    \label{Gamma}
\end{equation}
Agreement validating  Eq. \ref{Gamma} was reported for 12 even-even rotating nuclei with $R_{_{4/2}} > $ 2.7~\cite{allmond2017empirical}.
The same evaluation for all nuclei with $R_{_{4/2}} > $ 2.4, where  experimental data on four {\sc E2} matrix elements are available is shown in Fig.~\ref{fig-Gamma-vs-gamma}.
Discrepancies are observed for $^{76}$Kr and $^{194,196}$Pt, probably arising from the  mixed nature of the corresponding 2$^+_{_\gamma}$ states.
Indeed, shape coexistence at low-excitation energy was confirmed in $^{76}$Kr, while the observed 2$^+_{_\gamma}$ states in the two Pt isotopes are observed to decay through transitions with {\sc E0} components~\cite{wood1999electric}, which also suggests the presence of co-existing shapes~\cite{heyde2011shape}.

Henceforth, we extend the application of the generalized {\sc TR} approach \cite{allmond2017empirical} by adopting Eq. \ref{Gamma} not as an assumption, but as an empirically established dependence. This allows the application of this model-independent approach to a much larger range of rotating nuclei.

Specifically, from  Eqs.~\ref{ME01} and \ref{ME21}  we define the ratio $R_{_{22/02}}$ that is based on the two typically 
well-known
\linebreak $\langle 2^+_{_1} \parallel \hat{E2} \parallel 2^+_{_1}\rangle$ and $\langle0^+_{_1} \parallel \hat{E2} \parallel 2^+_{_1}\rangle$ matrix elements,
\begin{equation}
R_{_{22/02}} \coloneqq \frac{\langle2^+_{_1} \parallel \hat{E2} \parallel 2^+_{_1}\rangle } { \langle0^+_{_1} \parallel \hat{E2} \parallel 2^+_{_1}\rangle} = -\sqrt{\frac{10}{7}} \frac{\cos(\gamma -2\Gamma)}{\cos(\gamma+\Gamma)},
\label{ratio1}
\end{equation}
which taking Eq. \ref{Gamma} into account becomes
\begin{equation}
 R_{_{22/02}}(\gamma) = -\sqrt{\frac{10}{7}}~ \dfrac{\cos\left(\gamma + \cos^{-1}\left(\frac{\cos\left(4\gamma\right)+2\cos\left(2\gamma\right)}{\sqrt{4\cos\left(6\gamma\right)+5}}\right) \right)}
 {\cos \left( \gamma - \frac{1}{2} \cos^{-1}\left(\frac{\cos\left(4\gamma\right)+2\cos\left(2\gamma\right)}{\sqrt{8\cos^2\left(3\gamma\right)+1}}\right) \right)}.
\end{equation}

\begin{figure*}[!ht]
\begin{center}
\includegraphics[width=7.5cm,height=6.cm,angle=-0]{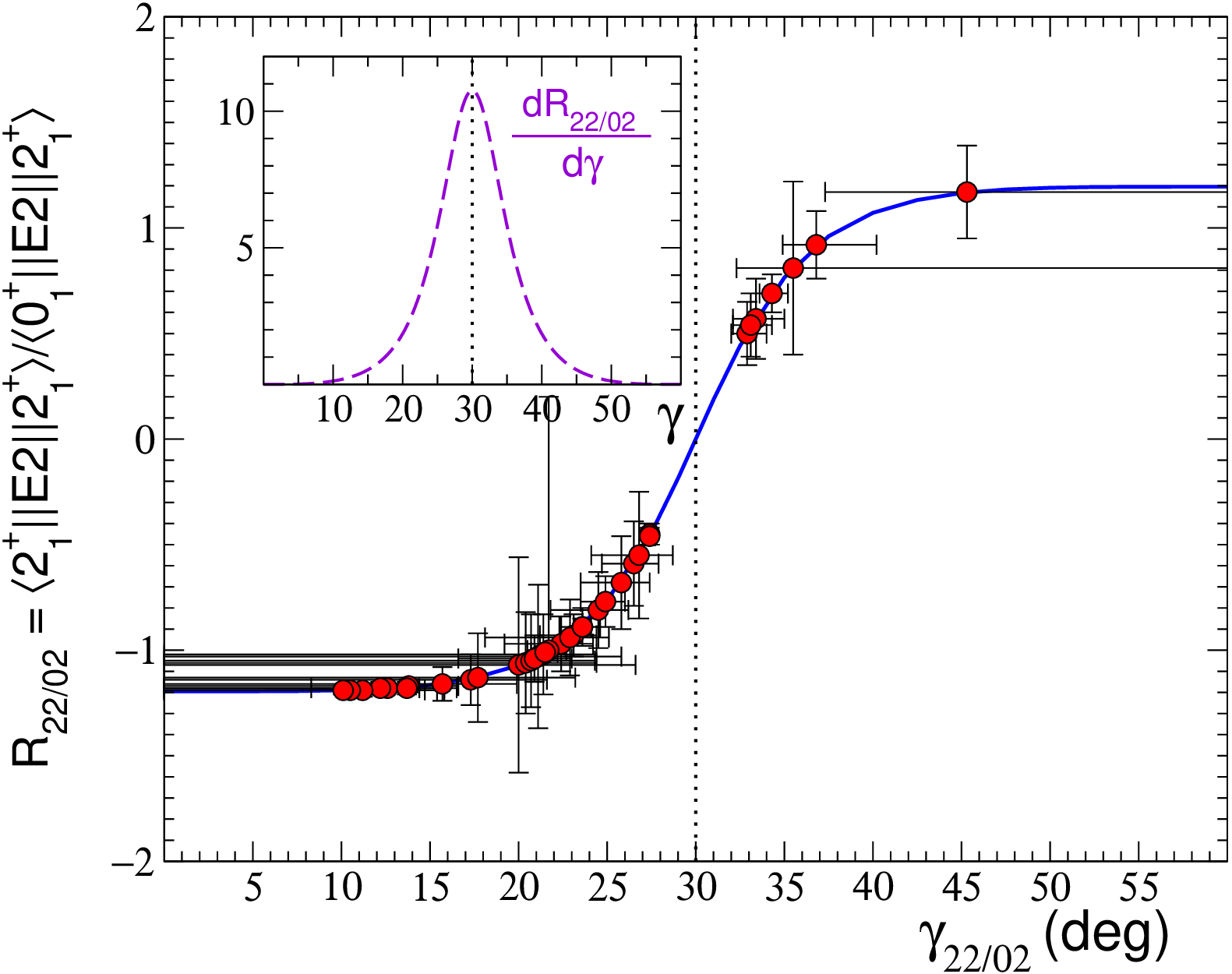}
\hspace{0.2cm}
\includegraphics[width=7.5cm,height=6.cm,angle=-0]{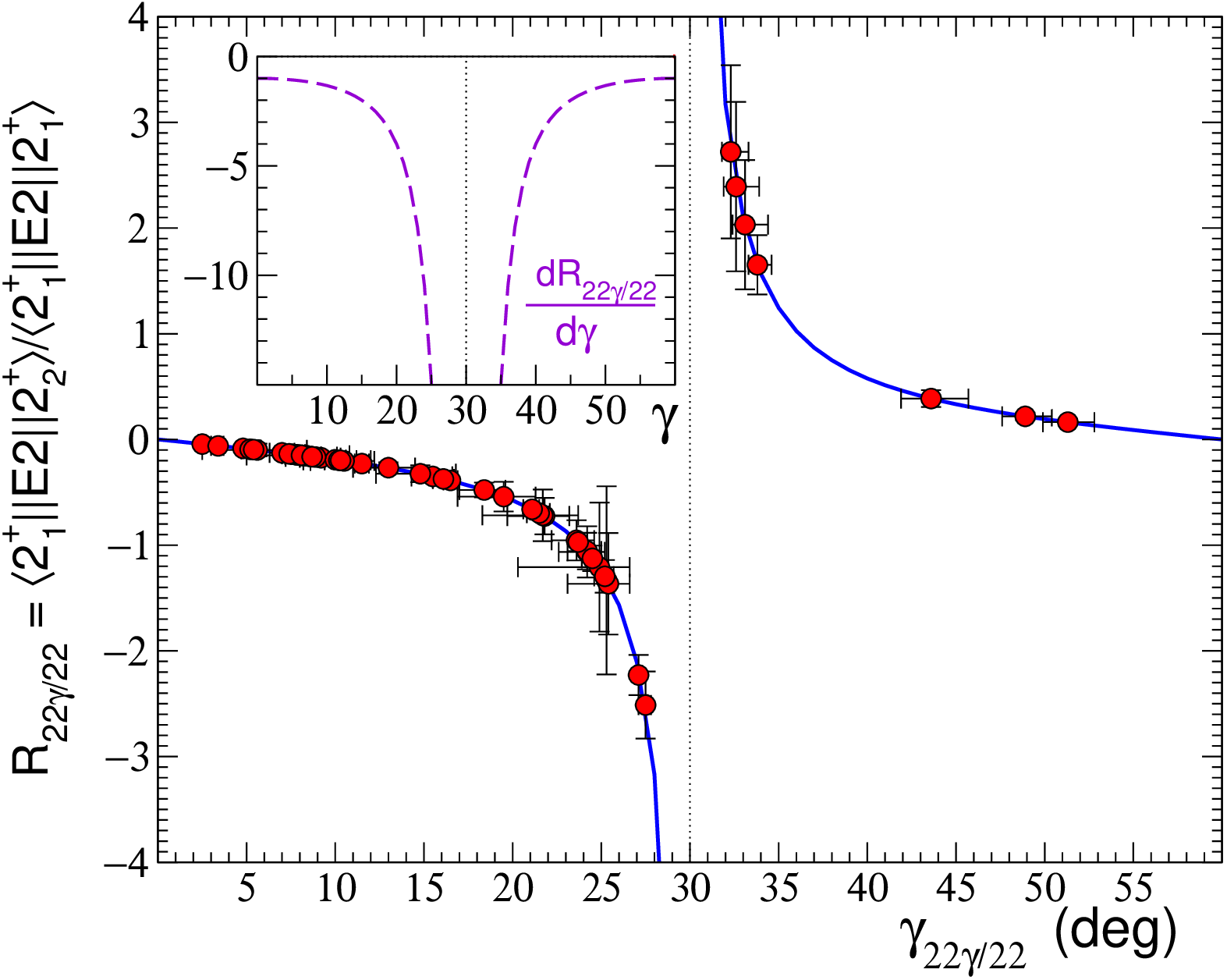}
\caption{$R_{_{22/02}}$ (left) and $R_{_{22\gamma/02}}$ (right)  ratios  as a function of the $\gamma$ deformation.
The solid curves are calculated within the proposed here approach while the experimental data (circles) are calculated from available matrix elements for even-even nuclei with $R_{_{4/2}}>2.4$. The corresponding $\gamma$ values are extracted from the theoretical (solid line) curves.
The first derivatives (dashed lines) reveal high sensitivity to the $\gamma$ degree of freedom.
}
\label{ratios}
\end{center}
\end{figure*}

The function $R_{_{22/02}}(\gamma)$, shown in the left panel of Fig. \ref{ratios} (solid line), is continuous in the $[0^{\circ},60^{\circ}]$ $\gamma$ range, varying smoothly between $R_{_{22/02}}(\gamma=0^{\circ})=-1.195$ (prolate) and $R_{_{22/02}}(\gamma=60^{\circ})=+1.195$ (oblate) and vanishing for $\gamma$ = 30$^\circ$. Therefore, one can deduce the $\gamma_{_{R22/02}}$ deformation of an even-even rotating nucleus using the $R_{_{22/02}}$ curve together with  $R_{_{22/02}}$ determined from the experimentally-determined matrix elements.
The first derivative $\frac{dR_{_{22/02}}}{d\gamma}$ (dashed line) shown in the inset of Fig. \ref{ratios} is also continuous, with a maximum at $\gamma=30^{\circ}$.
This method allows the precise determination of $\gamma$ values for nuclei with large asymmetry,
while considerable uncertainties are expected for nuclei with nearly axially-symmetric shapes.

Following this approach, we have evaluated $R_{_{22/02}}$ and $\gamma_{_{R22/02}}$ for 63 deformed even-even nuclei with $R_{4/2} >$ 2.4. These values are  shown in the left panel of Fig.~\ref{ratios} and listed in Table \ref{Table-data}  along with the corresponding
$\langle2^+_{_1} \parallel \hat{E2} \parallel 2^+_{_1} \rangle $ and  $\langle0^+_{_1} \parallel \hat{E2} \parallel 2^+_{_1}\rangle$ matrix elements, which were deduced from the corresponding evaluations of $Q_{_S} (2^+_{_1})$~\cite{stone2016table} and $B(E2; 0^+_{_1}\rightarrow 2^+_{_1})$~\cite{pritychenko2016tables} values, respectively, unless more recent and 
precise experimental data were available.
Deformations are labelled as prolate or oblate in the few cases where  $|R_{_{22/02}}|$ > 1.195, depending on the sign of $R_{_{22/02}}$.
There was no sign measured for the $\langle2^+_{_1} \parallel \hat{E2} \parallel 2^+_{_1} \rangle $ matrix element of $^{160}$Dy, we adopted negative sign in agreement with the systematics, see note {\sc p} in Table 1. The sign of the $\langle2^+_{_1} \parallel \hat{E2} \parallel 2^+_{_\gamma} \rangle $ matrix element in $^{76}$Kr was changed to positive (together with the same change for the $\langle2^+_{_\gamma} \parallel \hat{E2} \parallel 0^+_{_1} \rangle $ matrix element, which keeps the sign of the $P_3$ term unchanged~\cite{allmond2009triaxial}), in order to comply with the systematically observed prolate-type shapes of the neighbouring nuclei. The $P_3$ term is defined in Coulomb-excitation theory~\cite{kumar1969signs} as the interference between the direct excitation amplitude $0{_1}^+ \rightarrow 2_{_1}^+$ and the indirect one, $0_{_1}^+ \rightarrow 2^+_{_\gamma} \rightarrow 2_{_1}^+$, and depends on the product of the three related matrix elements;
Refs.~\cite{clement2007shape,rhodes2022evolution}).
One of the nice achievements of the generalized {\sc TR} model is that it can explain the sign of the $P_{_3}$ term~\cite{allmond2009triaxial}.

\begin{figure*}[!ht]
\begin{center}
\includegraphics[width=6.5cm,height=4.5cm,angle=-0]{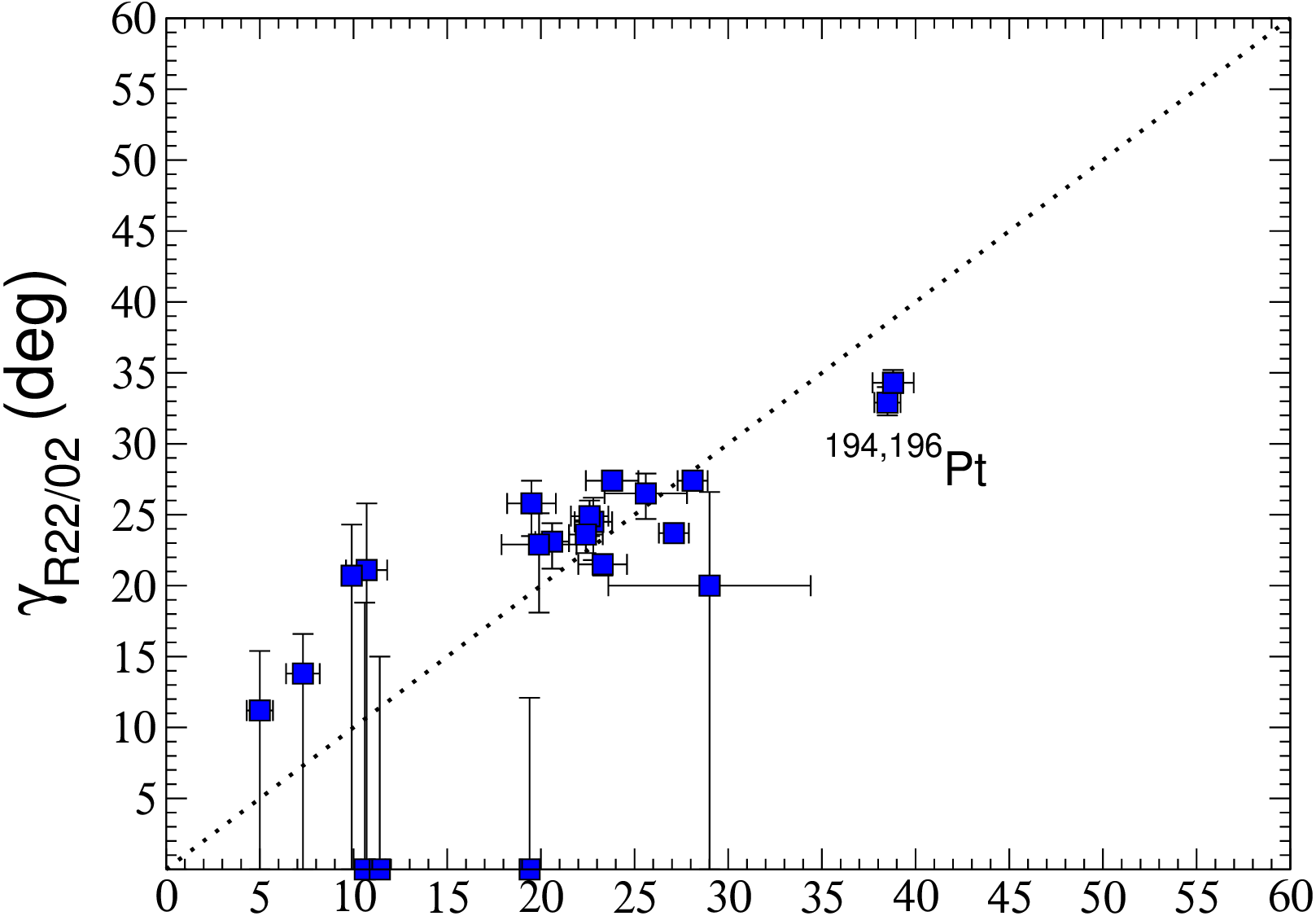} \hspace{0.3cm}
\includegraphics[width=6.2cm,height=4.5cm,angle=-0]{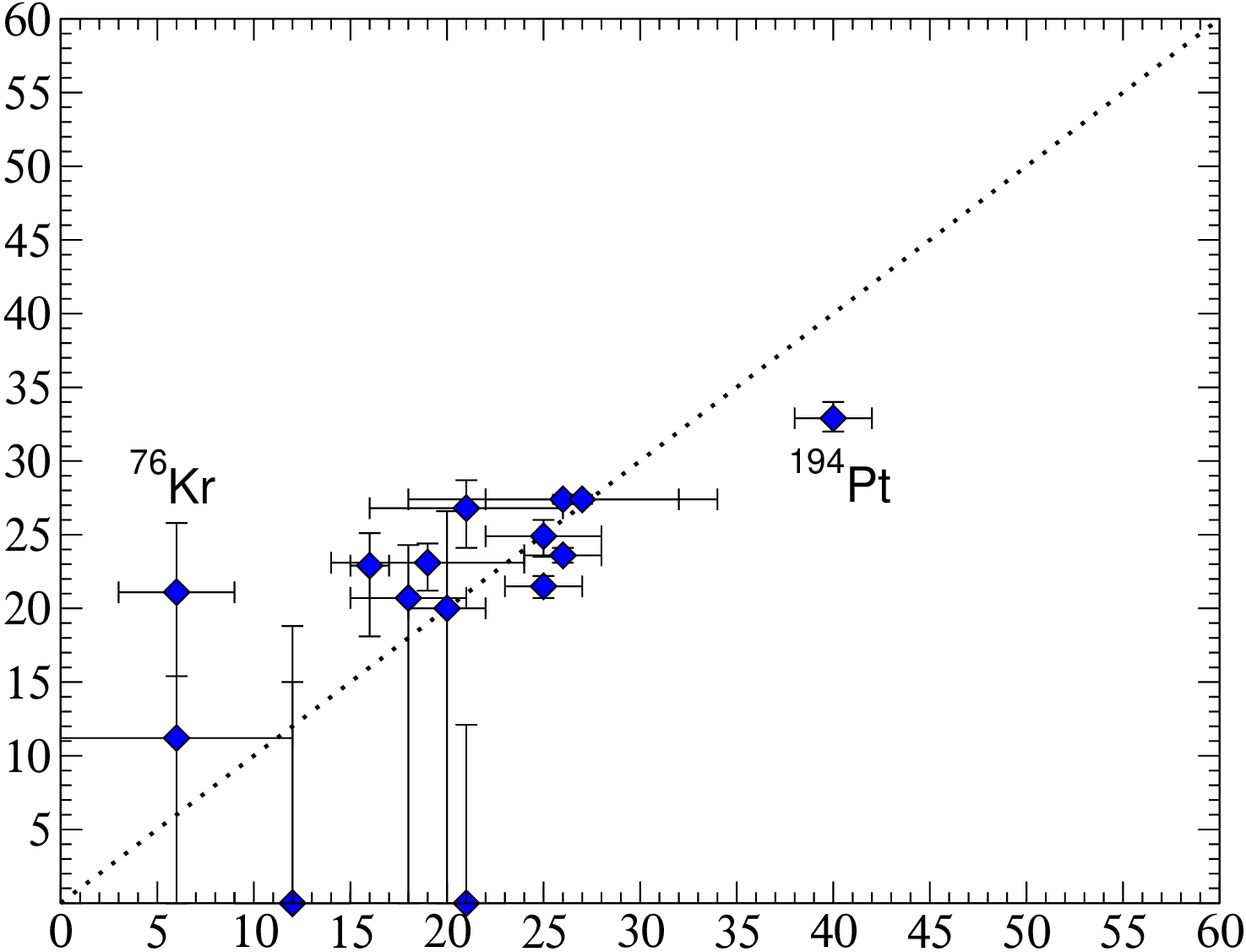}\\ \vspace{0.2cm}
\includegraphics[width=6.5cm,height=5.cm,angle=-0]{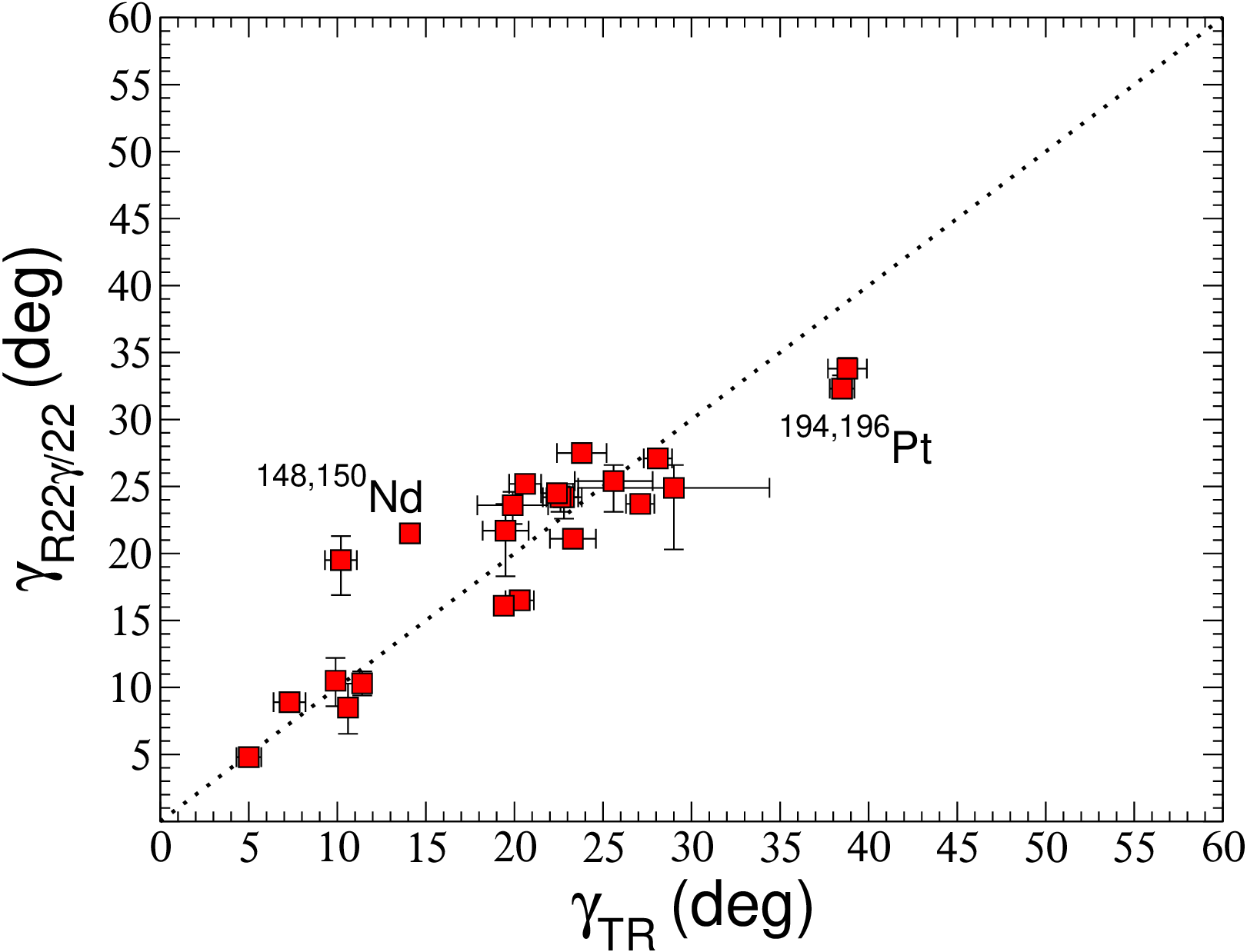} \hspace{0.3cm}
\includegraphics[width=6.2cm,height=5.cm,angle=-0]{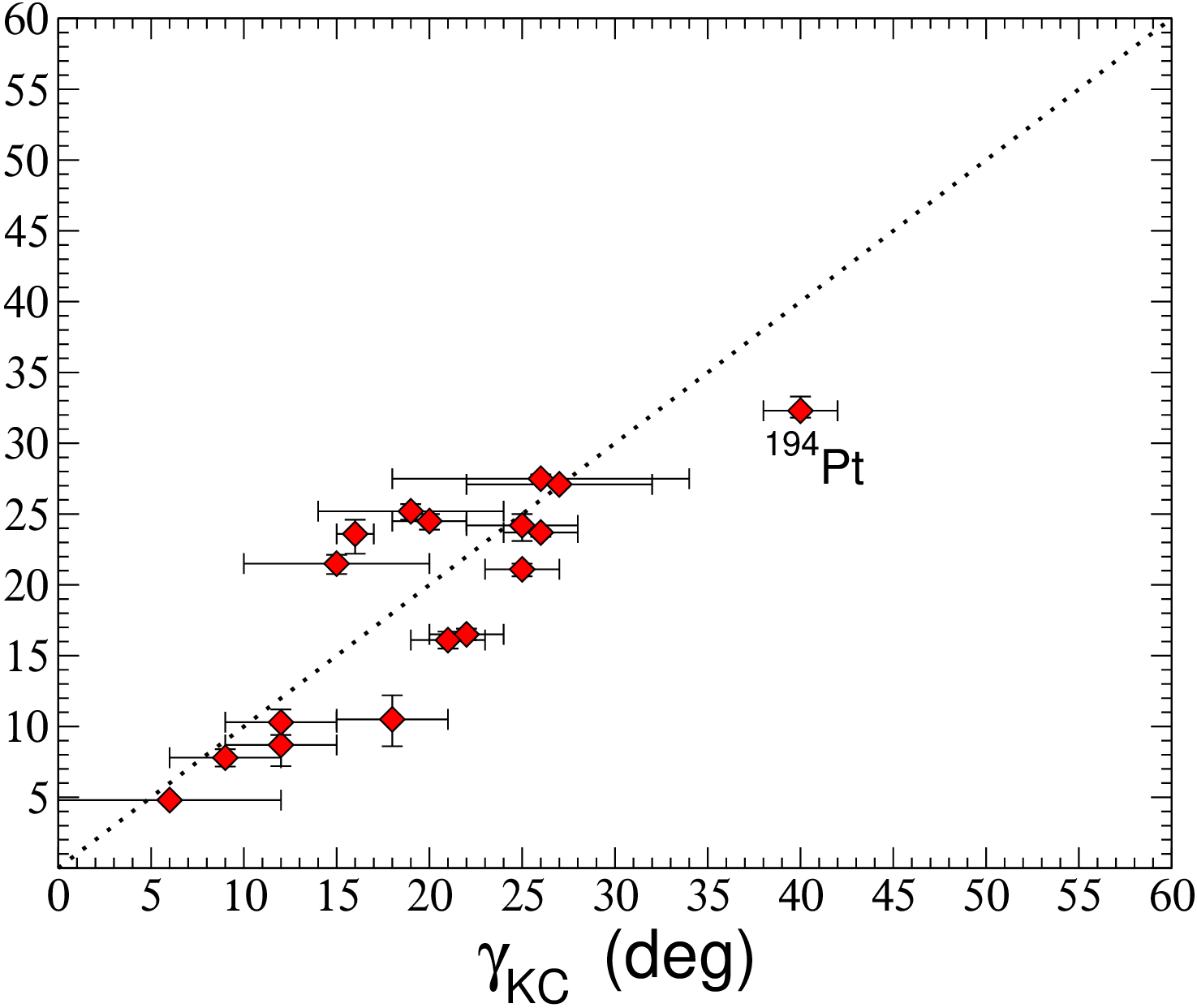}
\caption []{\label{fig:triaxialitiescomparison} Triaxial deformations $\gamma_{_{R_{_{22/02}}}}$ (top panels) and $\gamma_{_{R_{_{22\gamma/22}}}}$ (bottom panels)
versus $\gamma_{_{TR}}$  (left) and  $\gamma_{_{KC}}$ (right).}
\end{center}
\end{figure*}

For some of the nuclei analysed in our work there are previous assumptions-free evaluations of $\gamma_{_{TR}}$ and/or $\gamma_{_{KC}}$. A comparison of these values with those established in the proposed approach shows an overall agreement, with most values overlapping at the one- or two-$\sigma$ level, see the top panels of Fig. \ref{fig:triaxialitiescomparison}. Deviations are noticeable for the $^{194,196}$Pt nuclei where, as mentioned above, 
shape-coexisting effects are expected to play a role in the formation of the 2$^+_{_\gamma}$ states.

Thus, the $\gamma$ deformations were evaluated in an assumptions-free method for thirty even-even rotating nuclei beyond those for which $\gamma_{_{TR}}$ and $\gamma_{_{KC}}$ were available.
Many of these nuclei were found consistent with small triaxial deformations, not excluding axial symmetry, 
but we also identified a considerable number of triaxial nuclei, including
$^{56}$Fe ($\gamma_{_{R22/02}}$ = 22.4$^{+1.8}_{-3.2}$), 
$^{78}$Kr ($\gamma_{_{R22/02}}$ = 21.7$^{+1.0}_{-1.3}$), 
$^{152}$Sm ($\gamma_{_{R22/02}}$ = 12.6$^{+1.8}_{-4.3}$),
$^{170}$Er ($\gamma_{_{R22/02}}$ = 20.9$^{+2.1}_{-4.4}$),
$^{192}$Pt ($\gamma_{_{R22/02}}$ = 33.4$^{+1.6}_{-1.3}$),
$^{198}$Pt ($\gamma_{_{R22/02}}$ = 33.2$^{+1.2}_{-1.0}$),
and $^{198}$Hg ($\gamma_{_{R22/02}}$ = 36.8$^{+3.4}_{-1.9}$).
It should also be noted that $\gamma$ values inferred from this analysis are fully independent of the presence and features of the 2$^+_{_\gamma}$ $\gamma$ band. For instance, it  allows the assignment of triaxiality for nuclei where the $\gamma$ band  has not yet been established, as we did for $^{56}$Fe.
In addition, it permits an evaluation of triaxiality for nuclei where the 2$^+_{_\gamma}$ band is competing with other shape-coexisting structures and is, therefore, mixed. For instance, we have inferred triaxiality for $^{78}$Kr, 
$^{152}$Sm, and $^{170}$Er, where strong shape-coexisting phenomena occur~\cite{becker2006coulomb,park2016shape,kulp2007shape,heyde2011shape}, based entirely on the matrix elements of their 2$^+_1$ states. We have also established triaxial deformations for the $^{192,198}$Pt isotopes that are in agreement with the systematics (similar to the available $\gamma_{_{KC}}$ value of the neighbouring $^{196}$Pt isotope~\cite{wu1996quadrupole,lim1992measurements}) and proposed a triaxial shape for $^{198}$Hg, in contrast to the common assumption that the heavy Hg isotopes have axially-symmetric oblate deformations.
More details about the structural implications of these results will be presented in a separate manuscript.


\begin{table*}[!ht]
\centering
\caption{\label{Table-data}\small{The transitional and diagonal matrix elements (in units of eb) used to calculate the $R_{_{22/02}}$  and $R_{_{22\gamma/02}}$ ratios and the extracted $\gamma$ deformation based on these ratios. For comparison the $\gamma$ deformations deduced wherever possible using four matrix elements and using the 
Kumar-Cline rule are also listed. The data for the diagonal
$\langle2^+_{_1}\parallel \hat{E2} \parallel2^+_{_1}\rangle$ as well as the transitional $\langle 0^+_{_1}\parallel \hat{E2} \parallel 2^+_{_1}\rangle$  and $\langle 2^+_{_1} \parallel \hat{E2} \parallel2^+_{_\gamma}\rangle$ matrix elements are taken from Refs. \cite{stone2016table,pritychenko2016tables,pritychenko2012national}, unless stated differently. }}
\begin{tabular}{|cccccccc|}
\hline \hline
Nucleus &  $\langle2^+_{_1}\parallel \hat{E2} \parallel2^+_{_1}\rangle$   &  $\langle 0^+_{_1}\parallel \hat{E2} \parallel 2^+_{_1}\rangle$   &   $\langle 2^+_{_1} \parallel \hat{E2} \parallel2^+_{_\gamma}\rangle$  & $\gamma_{_{R22/02}}$     &  $\gamma_{_{R22\gamma/22}}$	  & $\gamma_{_{TR}}$  &  $\gamma_{_{KC}}$ \\
\hline
$^{12}$C & 0.125(24)$^{a}$ &  0.063(2) &       &  oblate &        &     &     \\
$^{20}$Ne & -0.303(40) &  0.182(4) &   0.052(3) &  prolate &  9.2$^{+1.1 }_{-1.2 }$ &     &     \\
$^{22}$Ne & -0.284(16)$^{b}$&  0.152(1) &   0.043(17) &  prolate &  8.1$^{+2.7 }_{-3.1 }$ &     &     \\
$^{22}$Mg & -0.57(57)$^{b}$ &  0.184(43) &      &  0$^{+30.5}_{-0 }$ &        &     &     \\
$^{24}$Mg & -0.237(26)$^{c}$ &  0.209(2) &   0.083(3) &  17.3$^{+4.3 }_{-17.3 }$ &  15.5$^{+1.1 }_{-1.2 }$ &     &     \\
$^{28}$Si	&	0.211(40) &	 0.181(2) &	 &	45.3$^{+14.7}_{-8.0}$ &	& & \\
$^{50}$Cr & -0.475(92) &  0.324(5) &   &  0$^{+13.1 }_{-0 }$ &   &     &     \\
$^{56}$Fe & -0.303(40) &  0.313(3) &   0.145(9) &  22.4$^{+1.8 }_{-3.2 }$ &  18.4$^{+1.2 }_{-1.4 }$ &     &     \\
$^{58}$Fe & -0.356(66) &  0.349(9) &   0.258(39) &  21.4$^{+3.0 }_{-21.4 }$ &  21.8$^{+1.4 }_{-2.1 }$ &     &     \\
$^{62}$Fe & -0.11(53)$^{d}$ &  0.319(97) &      &  28.2$^{+31.8 }_{-28.2 }$ &        &     &     \\
$^{74}$Ge & -0.251(26) &  0.553(14) &   0.630(44) &  27.4$^{+0.3 }_{-0.3 }$ &  27.5$^{+0.3 }_{-0.4 }$ & 23.8(14) & 26(8) \\
$^{76}$Ge & -0.240(20)$^{e}$ &  0.526(20)$^{e}$ &   0.535(7)$^{e}$ &  27.3$^{+0.3 }_{-0.3 }$ &  27.1$^{+0.2 }_{-0.2 }$ & 28.1(8) & 27(5) \\
$^{80}$Ge & -0.61(41)$^{f}$ &  0.408(10)$^{f}$ & $<|0.8|^{f}$    &  0$^{+27.2 }_{-0 }$ &   $<$25.3 or $>$34.7     &     &     \\
$^{78}$Se & -0.34(12) &  0.586(10) &   0.469(19) &  26.5$^{+1.4 }_{-1.8 }$ &  25.4$^{+1.2 }_{-2.3 }$ & 25.6(22) &     \\
$^{80}$Se & -0.409(92) &  0.502(8) &   0.435(12) &  24.5$^{+1.7 }_{-2.7 }$ &  24.2$^{+1.0 }_{-1.6 }$ & 22.8(10) &     \\
$^{82}$Se & -0.290(92) &  0.428(12) &   0.208(25) &  25.8$^{+1.6 }_{-2.3 }$ &  21.7$^{+2.0 }_{-3.4 }$ & 19.5(13) &     \\
$^{76}$Kr & -0.9(3)$^{g}$ &  0.871(15) &   0.09(4)$^{g,o}$ &  21.1$^{+4.7 }_{-21.1 }$ &  5.6$^{+2.8 }_{-3.1 }$ &  10.7(1.1)   &  6(3)   \\
$^{78}$Kr & -0.80(4)$^{h}$ &  0.796(10) &   0.26(6)$^{h}$ &  21.7$^{+1.0 }_{-1.2 }$ &  14.8$^{+2.0 }_{-2.5 }$ &     &     \\
$^{98}$Sr & -0.63(32)$^{i}$ &  1.14(20) &   &  26.8$^{+1.9 }_{-2.7 }$ &  &     &  21(3)   \\
$^{104}$Ru & -0.71(11)$^{j}$  &  0.917(25)$^{j}$ &   0.75(4)$^{j}$ &  24.9$^{+1.1 }_{-1.4 }$ &  24.2$^{+0.8 }_{-1.1 }$ & 22.6(10) & 25(3) \\
$^{110}$Ru & -1.10(52)$^{k}$ &  1.022(37)$^{k}$ &   1.32(25)$^{k}$ &  20.0$^{+6.6 }_{-20.0 }$ &  24.9$^{+1.7 }_{-4.6 }$ & 29.0(54) &     \\
$^{106}$Pd & -0.72(7)$^{l}$ &  0.812(10) &   0.810(37)$^{s}$ &  23.6$^{+1.0 }_{-1.3 }$ &  24.5$^{+0.5 }_{-0.6 }$ & 22.4(9) & 20(2) \\
$^{108}$Pd & -0.810(90)$^{l}$ &  0.874(11) &   1.049(44) &  23.1$^{+1.3 }_{-1.9 }$ &  25.2$^{+0.5 }_{-0.6 }$ & 20.6(9) & 19(5) \\
$^{110}$Pd & -0.87(17)$^{m}$ &  0.930(12) &   0.830(28) &  22.9$^{+2.2 }_{-4.8 }$ &  23.6$^{+1.0 }_{-1.4 }$ & 19.9(20) & 16(1) \\
$^{130}$Ba & -1.35(20) &  1.067(22) &      &  0$^{+19.9 }_{-0 }$ &        &       &     \\
$^{148}$Nd & -1.93(18) &  1.157(13) &   1.342(17) &  prolate &  21.5$^{+0.6 }_{-0.7 }$ & 14.1(3) & 15(5) \\
$^{150}$Nd & -2.64(66) &  1.645(9) &   1.427(9) &  prolate &  19.5$^{+1.8 }_{-2.6 }$ & 10.2(9) &     \\
$^{152}$Sm & -2.198(21) &  1.860(1) &   0.422(29) &  12.6$^{+1.8 }_{-4.3 }$ &  10$^{+0.6 }_{-0.6 }$ &       &     \\
$^{154}$Sm & -2.467(53) &  2.084(11) &  0.108(8)  &  12.2$^{+3.6 }_{-12.2 }$ & 2.5$^{+0.2}_{-0.2}$  &       &     \\
$^{154}$Gd & -2.401(53) &  1.968(4) &   0.549(22)$^{s}$ &  0$^{+7.9 }_{-0 }$ &  11.5$^{+0.4 }_{-0.4 }$ &       &     \\
$^{156}$Gd & -2.546(53) &  2.168(25) &   0.425(7) &  13.8$^{+2.8 }_{-13.8 }$ &  8.9$^{+0.2 }_{-0.2 }$ & 7.3(9) &     \\
$^{158}$Gd & -2.652(53) &  2.256(24) &   0.390(23) &  13.7$^{+2.8 }_{-13.7 }$ &  7.9$^{+0.4 }_{-0.4 }$ &       &     \\
$^{160}$Gd & -2.744(53) &  2.277(3) &   0.166(16) &  0$^{+12.5 }_{-0 }$ &  3.43$^{+0.3 }_{-0.4 }$ &       &     \\
$^{160}$Dy & -2.38(53)$^{p}$ &  2.247(9) &   0.468(17) &  20.4$^{+4.0 }_{-20.4 }$ &  10.2$^{+1.8 }_{-2.0 }$ &       &     \\
$^{164}$Dy & -2.74(20) &  2.370(14) &   0.444(18) &  15.7$^{+4.2 }_{-15.7 }$ &  8.6$^{+0.6 }_{-0.6 }$ &       &     \\
$^{166}$Er & -2.51(53) &  2.397(19) &   0.510(16) &  20.7$^{+3.6 }_{-20.7 }$ &  10.5$^{+1.7 }_{-1.9 }$ & 9.9(5) & 18(3) \\
$^{168}$Er & -3.25(25)$^{n}$  &  2.43(7)$^{n}$  & 0.47(2)$^{n}$  &  prolate &  7.8$^{+0.6 }_{-0.6 }$ & 8.2(3) & 9(3) \\
$^{170}$Er & -2.51(27) &  2.416(14) &  $>$ 0.385 &  20.9$^{+2.1 }_{-4.3 }$ &  $>$ 8.3 &       &     \\
$^{170}$Yb & -2.876(40) &  2.392(15) &   0.366(38)$^{r}$ &  0$^{+12 }_{-0 }$ &  7.0$^{+0.7 }_{-0.7 }$ &       &     \\
$^{172}$Yb & -2.929(53) &  2.468(30) &   0.250(6) &  11.2$^{+4.2}_{-11.2}$ &  4.8$^{+0.1}_{-0.1}$ & 5.0(7) & 6 (6) \\
$^{174}$Yb & -2.876(66) &  2.419 (33) & 0.269(27)$^{r}$ &  10.5$^{+5.3}_{-10.5 }$ &  5.2 $^{+0.5}_{-0.5}$ &  &     \\
$^{176}$Yb & -3.008(79) &  2.278(20) &  0.289(19) &  prolate &  5.4$^{+0.4 }_{-0.4 }$ &  &     \\
$^{176}$Hf & -2.771(26) &  2.328(37) & 0.387(36)$^{r}$   &  10.1$^{+4.6 }_{-10.1 }$ & 7.6$^{+0.6}_{-0.6}$ &       &     \\
$^{178}$Hf & -2.665(26) &  2.176(145) &   0.362(12) &  0$^{+16.9 }_{-0 }$ &  7.4$^{+0.2 }_{-0.2 }$ &       &     \\
\hline
 \end{tabular}
\end{table*}

\begin{table*}
\centering
\begin{tabular}{|cccccccc|}
\hline \hline
Nucleus &  $\langle2^+_{_1}\parallel \hat{E2} \parallel2^+_{_1}\rangle$   &  $\langle 0^+_{_1}\parallel \hat{E2} \parallel 2^+_{_1}\rangle$   &   $\langle 2^+_{_1} \parallel \hat{E2} \parallel2^+_{_\gamma}\rangle$  & $\gamma_{_{R22/02}}$     &  $\gamma_{_{R22\gamma/22}}$	  & $\gamma_{_{TR}}$  &  $\gamma_{_{KC}}$ \\ \hline
$^{180}$Hf & -2.639(26) &  2.156(1) &   0.396(23) &  prolate &  8.1$^{+0.4 }_{-0.4 }$ &       &     \\
$^{180}$W & -2.77(53) &  2.037(34) &      &  0$^{+19 }_{-0 }$ &        &       &     \\
$^{182}$W & -2.77(53) &  2.031(10) &   0.454(6)$^{s}$ &  0$^{+18.8 }_{-0 }$ &  8.7$^{+1.4 }_{-1.5 }$ & 10.6(2) & 12(3) \\
$^{184}$W & -2.51(27) &  1.925(9) &   0.497(7) &  0$^{+15 }_{-0 }$ &  10.3$^{+0.9 }_{-0.9 }$ & 11.4(3) & 12(3) \\
$^{186}$W & -2.11(40) &  1.871(10) &   0.564(20) &  17.7$^{+5.5 }_{-17.7 }$ &  13.0$^{+1.5 }_{-2.0 }$ &       &     \\
$^{184}$Os & -3.6(16) &  1.793(22) &      &  0$^{+18.9 }_{-0 }$ &        &       &     \\
$^{186}$Os & -2.151(53) &  1.750(21) &   0.835(32) &  prolate &  16.5$^{+0.4 }_{-0.4 }$ & 20.3(8) & 22(2) \\
$^{188}$Os & -1.926(53) &  1.581(11) &   0.720(40) &  0$^{+12.1 }_{-0 }$ &  16.1$^{+0.6 }_{-0.6 }$ & 19.4(5) & 21(2) \\
$^{190}$Os & -1.557(40) &  1.534(29) &   1.028(54) &  21.5$^{+0.7 }_{-0.8 }$ &  21.1$^{+0.4 }_{-0.5 }$ & 23.3(13) & 25(2) \\
$^{192}$Os & -1.267(40) &  1.425(35) &   1.230(35) &  23.6$^{+0.5 }_{-0.5 }$ &  23.7$^{+0.2 }_{-0.3 }$ & 27.1(8) & 26(2) \\
$^{192}$Pt & 0.79(27) &  1.393(23) &   1.894(61) &  33.4$^{+1.6 }_{-1.3 }$ &  32.6$^{+1.3}_{-0.7 }$ &       &     \\
$^{194}$Pt & 0.63(0.19) &  1.277(27) &   1.72(12)$^{s}$ &  32.9$^{+1.1 }_{-0.9 }$ &  32.3$^{+1.0 }_{-0.5 }$ & 38.5(7) & 40(2) \\
$^{196}$Pt & 0.82(0.11) &  1.184(29) &   1.35(15)$^{s}$ &  34.3$^{+0.9 }_{-0.7 }$ &  33.8$^{+0.8 }_{-0.5 }$ & 38.8(11) &     \\
$^{198}$Pt & 0.55(16) &  1.035(24) &   1.13(0.11) &  33.1$^{+1.2 }_{-1.0 }$ &  33.1$^{+1.3 }_{-0.7 }$ &     &     \\
$^{198}$Hg & 0.90(0.16) &  0.980(4) &   0.147(9) &  36.8$^{+3.4 }_{-1.9 }$ &  51.3$^{+1.5 }_{-1.4 }$ &     &     \\
$^{200}$Hg & 1.27(0.15) &  0.925(15) &   0.276(31) &  oblate &  48.9$^{+1.5 }_{-1.3 }$ &     &     \\
$^{202}$Hg & 1.15(0.18) &  0.784(13) &   0.444(59) &  oblate &  43.6$^{+2.1 }_{-1.7 }$ &     &     \\
$^{204}$Hg & 0.53(27) &  0.651(16) &      &  35.5$^{+24.5 }_{-3.2 }$ &        &     &     \\
\hline
 \end{tabular}

$^{a}$ from Ref. \cite{saiz2023spectroscopic};
$^{b}$ from Ref. \cite{henderson2018testing};
$^{c}$ from Ref. \cite{spear1981static};
$^{d}$ from Ref. \cite{gaffney2015low};
$^{e}$ from Ref. \cite{ayangeakaa2019evidence};
$^{f}$ from Ref. \cite{rhodes2022evolution};
$^{g}$ from Ref. \cite{clement2007shape};
$^{h}$ from Ref. \cite{becker2006coulomb};
$^{i}$ from Ref. \cite{clement2016spectroscopic};
$^{j}$ from Ref. \cite{srebrny2006experimental};
$^{k}$ from Ref. \cite{doherty2017triaxiality};
$^{l}$ from Ref. \cite{svensson1995multiphonon};
$^{m}$ from Ref. \cite{svensson1989coulomb};
$^{n}$ from Ref. \cite{kotlinski1990coulomb};
$^{o}$ the sign of the entry was changed, see text for more details 
$^{p}$ the entry has no sign, negative sign is assumed, see text for more details; 
$^{q}$ $\gamma$ = 60$^{\circ} - \gamma$ is also possible;
$^{r}$ pure {\sc E2} is assumed;
$^{s}$ the transition has {\sc E0} component.

\end{table*}

We have also defined another ratio of matrix elements, 
$R_{_{22\gamma/22}}$,  
\begin{equation}
R_{_{22\gamma/22}} \coloneqq \frac{\langle2^+_{_1} \parallel \hat{E2} \parallel 2^+_{_\gamma}\rangle } { \langle2^+_{_1} \parallel \hat{E2} \parallel 2^+_{_1}\rangle} = - \tan(\gamma - 2 \Gamma),
\label{ratio2}
\end{equation}
which allows to deduce nuclear triaxiality. 
Again, using Eq. \ref{Gamma},
\begin{equation}
R_{_{22\gamma/22}}(\gamma) = - \tan \left( \gamma + \cos^{-1} \left( \frac{\cos{(4\gamma)} + 2\cos{(2\gamma)}}{\sqrt{9-8\sin^2{(3\gamma)}}} \right) \right).
\end{equation}

The function $R_{_{22\gamma/22}}(\gamma)$ is shown in the right panel of Fig. \ref{ratios} (solid line), while its first derivative is plotted in the inset (dashed line).
This ratio has 
an advantage over $R_{_{22/02}}$ because of its sensitivity to $\gamma$  throughout the full $[0^\circ, 60^\circ]$ range.
It can be applied to all even-even rotating nuclei where the 2$^+_\gamma$ band head is well established, and not affected by shape co-existence or other phenomena. 
We have thus examined the available data for $\langle 2^+_{_1}\parallel\hat{E2}\parallel 2^+_{_\gamma}\rangle$ matrix elements in all deformed even-even rotating nuclei with $R_{4/2} > $ 2.4,
as listed in Table \ref{Table-data}.
In most cases the matrix element is deduced from the measured $B(E2;2^+_{_\gamma} \rightarrow 2^+_{_1})$ value.

In order to test this approach we have first calculated the $R_{_{22\gamma/22}}$ ratios and the corresponding $\gamma_{_{R22\gamma/22}}$ values for the deformed even-even nuclei where $\gamma_{_{TR}}$ and/or $\gamma_{_{KC}}$ are available.
Comparisons of $\gamma_{_{R22\gamma/22}}$ vs $\gamma_{_{TR}}$ and  $\gamma_{_{R22\gamma/22}}$ vs $\gamma_{_{KC}}$ are
shown in the bottom left and right panels of Fig. \ref{fig:triaxialitiescomparison}, respectively. Except for $^{148,150}$Nd and $^{194,196}$Pt, there is overall agreement between the axial asymmetries derived by these three different methods, most often within one or two $\sigma$ intervals. The discrepancies for the two Nd isotopes probably arise because of the presence of {\sc K = 0} excited bands lying at very similar excitation energy to the 2$^+_{_{\gamma}}$  bands (resulting in mixing of the 2$^+_{_\gamma}$ states), while the Pt isotopes were already discussed above. Thus, the agreement observed in Fig. \ref{fig:triaxialitiescomparison} validates the proposed determination of $\gamma$ based on the $R_{_{22\gamma/22}}$ ratio.
It should be noted that this method allows to determine $\gamma$ with good precision even for near axially-symmetric nuclei; for instance, $^{172}$Yb with $\gamma_{_{R22\gamma/22}}=4.8(1)^\circ$.
The $\gamma_{_{R22\gamma/22}}$ deformations
determined using the $R_{_{22\gamma/22}}$ ratios
are illustrated in the right panel of Fig. \ref{ratios} and listed in Tab.~\ref{Table-data} for 27 nuclei,
in addition to those previously determined through the $\gamma_{_{TR}}$ analysis~\cite{allmond2017empirical,23All-CWAN}.
The 
values of $\gamma$ derived from the $R_{_{22/02}}$ and $R_{_{22\gamma/22}}$ ratios 
are similar (except for $^{198}$Hg), 
and describe shapes with all possible triaxialities.

It is important to stress that $R_{_{22\gamma/22}}$ 
analysis assigned triaxial shapes to all the
53 even-even rotating nuclei where $\langle2^+_{_1} \parallel \hat{E2} \parallel 2^+_{_\gamma}\rangle $ and
$\langle2^+_{_1} \parallel \hat{E2} \parallel 2^+_{_1}\rangle $ are known.
This observation is in line with the 
suggestion that assumption-free analyses (such as
the model-independent $\gamma_{_{KC}}$ evaluation based on multi-step Coulomb-excitation measurements with sufficient statistics~\cite{wu1991electromagnetic,svensson1995multiphonon,wu1996quadrupole,srebrny2006experimental,ayangeakaa2019evidence}, and the generalized TR model),
establish triaxial deformations 
for the vast majority of the studied nuclei. 
These findings suggest that ideal axially-symmetric prolate or oblate nuclear rotors may not be 
common.


In summary, this work proposes the use of simple ratios,  
$R_{_{22/02}}$ and $R_{_{22\gamma/22}}$, of typically 
easy-to-measure {\sc E2} matrix elements  ($\langle0^+_{_1} \parallel \hat{E2} \parallel 2^+_{_1}\rangle$, $\langle2^+_{_1} \parallel \hat{E2} \parallel 2^+_{_1}\rangle$ and $\langle2^+_{_1} \parallel \hat{E2} \parallel 2^+_{_\gamma}\rangle $)
to extract the $\gamma$ deformation of 
even-even rotating nuclei in a model-independent way.
The approach is based on the Davidov-Filippov equations for the 2$^+_{_1}$ and 2$^+_{_\gamma}$ states of even-even rotating nuclei. It is parameter-free because all assumptions of the model 
were either proven empirically (irrotational-flow dependence of the MoI from $\gamma$) or become irrelevant (the spin dependence of the MoI).
It requires experimental data on two matrix elements only,  facilitating its application on a larger number of even-even rotating nuclei.
The $\gamma$ values determined 
using these ratios are in agreement with those established with the 
model-independent {\sc KC} sum rules approach and the generalized {\sc TR} model.
The $R_{_{22/02}}$ ratio analysis allows the precise identification of triaxial deformations in the range $20^\circ \lessapprox \gamma_{_{R22/02}} \lessapprox  40^\circ$
using the
{\sc E2} matrix elements of the 2$^+_{_1}$ state alone; hence, opening the interesting prospect of determining the triaxiality of exotic nuclei.
As this approach does not require knowledge of the 2$^+_{_\gamma}$ band, it is also very valuable for measuring triaxiality in nuclei where  shape coexistence appears at low excitation energies and affects the corresponding $\gamma$ band.
The $R_{_{22\gamma/22}}$ ratio analysis needs knowledge of the $\langle2^+_{_1} \parallel \hat{E2} \parallel 2^+_{_\gamma}\rangle $ matrix element and is very sensitive in the full 0$^\circ < \gamma_{_{R22\gamma/22}} < $ 60$^\circ$ range.
We report results from the proposed analyses on more than 60 even-even rotating nuclei where the axial asymmetries of the nuclear shapes are deduced in an assumption-free approach.

The work is based on research supported in part by the National Research Foundation of South Africa (Grant Number 150650).

\bibliographystyle{elsarticle-num}

\bibliography{triaxial}

\end{document}